\documentclass[twocolumn,showpacs,preprintnumbers,amsmath,amssymb]{revtex4}

\usepackage{graphicx}
\usepackage{dcolumn}
\usepackage{bm}

\begin{document}

\preprint{APS/123-QED}

\title{Transport and Magnetic properties of PrCoI$n_5$}
\author{Abebe Kebede}
\email{abkebede@gmail.com}
\homepage{http://sirius-c.ncat.edu/}
\author{Terrel Dial}
\affiliation{Department of Physics}
\address{North Carolina Agricultural and Technical State University}
\address{1601 East Market Street, Greensboro, NC 27411}
\date{\today}
\begin{abstract}
Structural, electrical and magnetic measurements of 115 single
crystals of PrIn$_5$ are reported. It has a tetragonal structure and
has slightly lower cell volume than its isomorphic counter part
CeCoIn$_5$. The resistivity saturates for T$\geq 10K$ . Analysis of
the resistivity for  10K$<$T$<$ 60$K$ indicates a regular fermi
liquid behavior. It does not exhibit superconductivity down to
T$\sim 1K$. The magnetic susceptibility analysis yielded the moment
to be 4.00$\mu_B$ indicating that the magnetism of PrCoIn$_5$ is
dominated by Pr$^{3+}$ free ions with some admixture of the magnetic
moment of the Co sublattice. The paramagnetic Curie temperature
$\theta \sim$-40$K$.  At low temperatures the susceptibility follows
a broad maximum around T$_N$ $\sim $ 14.5$K$, and increases as the
temperature is lowered. The disappearance of superconductivity for
T$>$1$K$ is attributed to chemical pressure effects and magnetic
pair breaking.
\end{abstract}

\pacs{74.70.Tx, 74.25.Bt, 74.62Bf}

\maketitle
\section{INTRODUCTION}
A typical metal is expected to exhibit fermi liquid behavior in that
at low temperatures the electronic specific heat coefficient,
$\gamma$ and the magnetic susceptibility, $\chi$ are constants, and
the resistivity, $\rho = \rho_0$ +AT$^2$. Instead  the normal state
properties of CeCoIn$_5$ are characterized by a non fermi liquid
behavior where $C/T$ $\sim -ln T$ and $\rho = \rho_0$+AT$^n$ ($n<
2$) and $\chi \sim T^{-n}$ ($n < 1$). Its superconducting transition
temperature T$_c$ $\sim$ 2.3$K$ is significantly influenced by
application of hydrostatic pressure and chemical substitution. In
the system CeMIn$_5$ (M=  Co, Rh, Ir), it has been shown that the
superconductivity can be induced by tuning Neel temperature, T$_N$
to zero.  Substitution of In with Sn\cite{Bauer} and Ce with
La\cite{Petrovic} produced pair breaking effects attributed to
chemical pressure\cite{Sparna}. In this work we synthesized good
quality PrCoIn$_5$ single crystals to study its structure, thermal
and magnetic properties. The structurally similar PrCoIn$_5$ is non
superconducting, its resistivity saturates below 10$K$.  These
differences in properties may alow researchers to study the
evolution from superconducting non fermi liquid grould state
exhibited by CeCoIn$_5$  to a non superconducting fermi liquid state
exhibited by PrCoIn$_5$. Single crystals of PrCoIn$_5$ were
synthesized from indium flux by combining stoichiometric amounts of
Pr and Co with excess indium. The charge is put inside an alumna
crucible, which is sealed in an evacuated quartz cylinder. It is
then heated to 1150 C at the rate of 2 C per minute, followed by
cooling to 750 C at 3C/minute, rapid cooling to 450C at 2C/minute.
The charge is then removed, and quickly put in centrifuge to remove
the excess flux. Each good crystal is etched with concentrated
hydrochloric acid for several hours and then it is rinsed with
ethanol  in ultrasonic environment. The resulting crystals are
mostly plates with dimensions $2mm\times 2mm\times 1mm $.
\begin{figure}[htb]
\center{\includegraphics[scale=0.25]{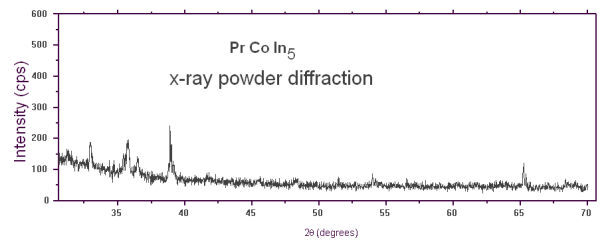}}
\caption{x-ray powder diffraction patterns of PrCoIn$_5$}
\label{fig:Fig}
\end{figure}
\begin{table}
\caption{Lattice parameters of PrCoIn$_5$}
\begin{ruledtabular}
\begin{tabular}{cccccc}
Sample& $a(\AA)$ & $b(\AA)$ & $c(\AA)$ & $c/a$ & Cell volume $(\AA^3)$\\
\hline
PrCoIn$_5$ & 4.61 & 4.61 & 7.50 & 1.62 & 161.09\\
\hline
CeCoIn$_5$ & 4.62 & 4.62 & 7.56 & 1.63 & 161.36\\
\hline
\end{tabular}
\end{ruledtabular}
\end{table}

The lattice parameters were obtained from single crystal x-ray diffraction pattern with Mo-K$\alpha $ radiation.  The x-ray intensity as a function of $2\theta$ is shwon in Figure 1 above.
The reflections were indexed for the tetragonal unit cell. Shown on Table 1 are the lattice parameters and the cell volumes of PrCoIn$_5$ and CeCoIn$_5$.  The $c$/$a$ $\sim$ 1.62 clearly indicates that the structure of PrCoIn$_5$ is teragonal and  it is stucturally isomorphic with CeCoIn$_5$.

\section{RESITIVITY  of PrCoIn$_5$}

Resistivity was measured using a standard 4- wire technique at
temperatures ranging from 300 K to 1.8 K in helium cooled cryostat.
Excitation currents of $500\mu A -1000\mu A$ were applied parallel
to the c-axis at frequencies of 16 Hz. Each sample was mounted on a
resistivity puck with GE varnish. $\#$40 guage copper wires  were
attached to the samples with silver epoxy. The magnetic
susceptibility and magnetization measurements were carried out with
a PPMS system at NHMFL-Los Alamos.

\begin{figure}[htb]
\center{\includegraphics[scale=0.60]{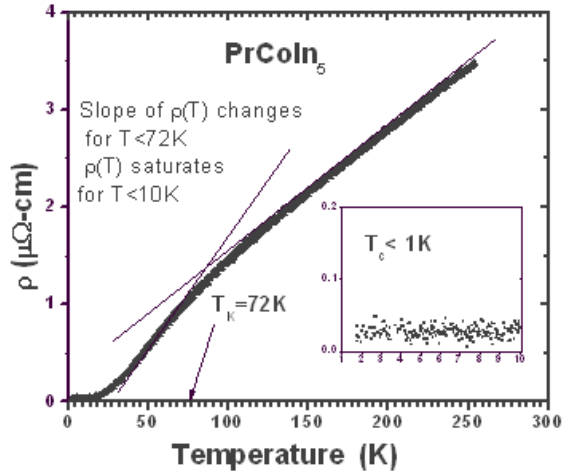}}
\caption{Reistivity versus Temperature of PrCoIn$_5$}
\label{fig:fig2}
\end{figure}

As shown in Figure 2  the resistivity shows metallic behavior above
about 150K. Since for typical metals such as copper, the deviation
from linearity begins at $T \sim 0.7\theta_D$. We estimated
$\theta_D \sim 241K$. Below $150K$ the data follow a broad maximum
near T $\sim 72K$ accompanied with slow decrease as the temperature
is lowered. The broad maximum is attributed to crystalline electric
field effects. There is no evidence of superconductivity within the
accessible temperature $T > 1K$. Shown in Figure 3 is our fit of the
data for $10K<T<60K$ using a polynomial  of order five. Assuming
that the contribution of the higher order terms diminishes as the
temperature is lowered, we retained the fit parameters, $\alpha
=0.034 \mu \Omega$-cm, $\beta=-0.005 \mu\Omega -cm/K$ and
$A=4.94\times 10^{-4} \mu\Omega-cm/K^2$. While the negative linear
term can be an artifact of the fit, the $T^2$ term is suggestive of
a normal fermi liquid behavior. Extrapolated from the high
temperature linear component we obtain the residual resistivity
$\rho_0 \sim 0.021 \mu \Omega-cm$ and the residual resistivity ratio
($RRR=\rho_{RT}/\rho_0$) was 200. This ratio indicates that this
crystal is of reasonable quality. As shown in Figure 4 the
resistivity saturates for $T< 10K$ with $\rho_{sat} \sim 0.034 \mu
\Omega-cm. \rho_{sat}$ is lower than the residual resistivity,
$\rho_o$. Whether or not the origin of this saturation is of
fundamental nature or caused by inclusions and surface effects is to
be determined.

\begin{figure}[htb]
\center{\includegraphics[scale=0.25]{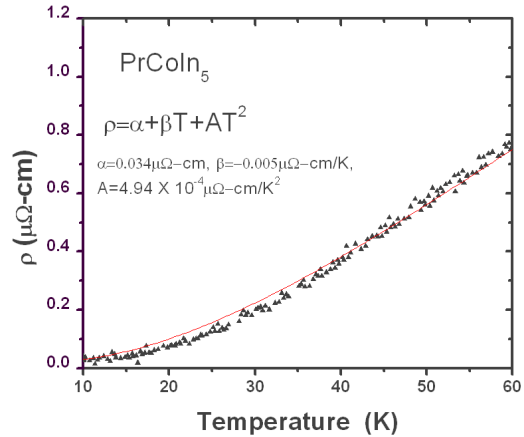}}
\caption{Low Temperature Resistivity of PrCoIn$_5$}
\label{fig:fig2}
\end{figure}

\begin{figure}[htb]
\center{\includegraphics[scale=0.25]{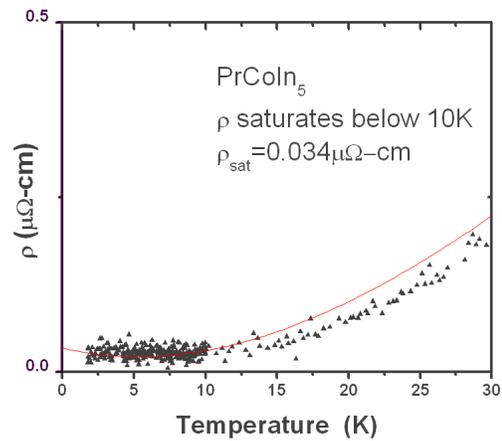}}
\caption{Saturation of the Resistivity versus Tememrature of PrCoIn$_5$}
\label{fig:fig2}
\end{figure}

\begin{table}
\caption{Physical parameters of PrCoIN$_5$}
\begin{ruledtabular}
\begin{tabular}{cccccc}
PrCoIn$_5$ & $T_K(K)$ & $T_c(K)$ & $T_N$ & $\theta(K)$ & $\mu(\mu_B)$\\
\hline
& 72 & 0 &14.5K& -40 & 4.00\\
\end{tabular}
\end{ruledtabular}
\end{table}

\section{Magnetic Susceptibility of PrCoIn$_5$}

Shown in Figure 5 is the magnetic susceptibility of PrCoIn$_5$ measured at 1kOe.  The data showed a rapid upturn at low temperature. The inset shows the susceptibtibility below 30K, the broad maximum near T$_N$$\sim$ 14.5K. This structure coincides with the resistivity saturation for T$<$ 10$K$ and we believe it is associated with antiferromagnetic ordering of the Pr ions. While similar behaviour has observed in other Pr based alloys such as PrBa$_2$Cu$_3$O$_7$, a non superconducting cuprate\cite{Kebede} and Pr$_3$In\cite{Christianson} the presence of multiphases cannot be ruled out. Shown in Figure 6 is the plot of the inverse susceptibility, $\chi^{-1}$ versus temerature.

\begin{figure}[htb]
\center{\includegraphics[scale=0.30]{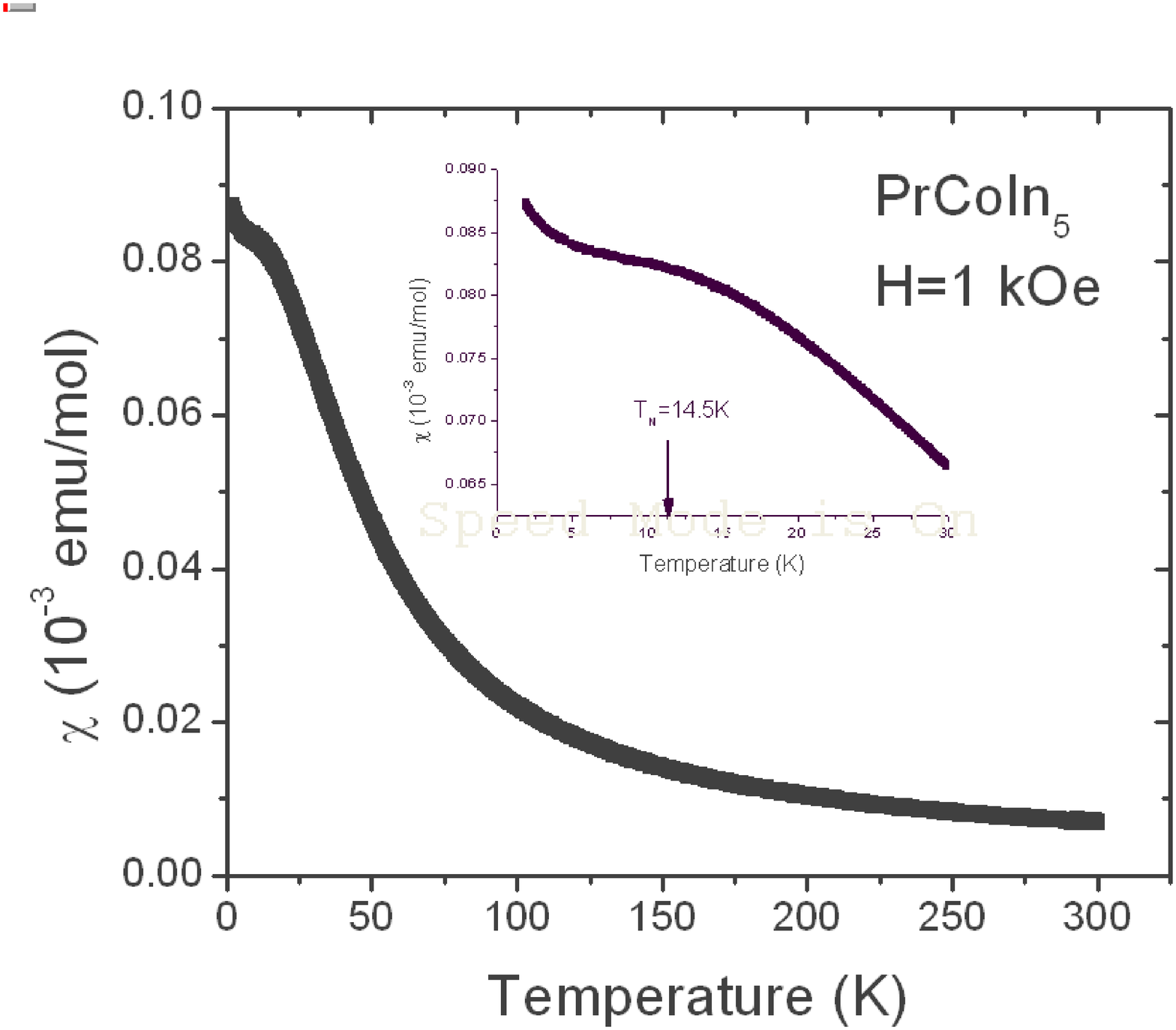}}
\caption{Magnetic Susceptibility versus Temperature of PrCoIn$_5$ at H=1kOe}
\label{fig:fig2}
\end{figure}
\begin{figure}[htb]
\center{\includegraphics[scale=0.30]{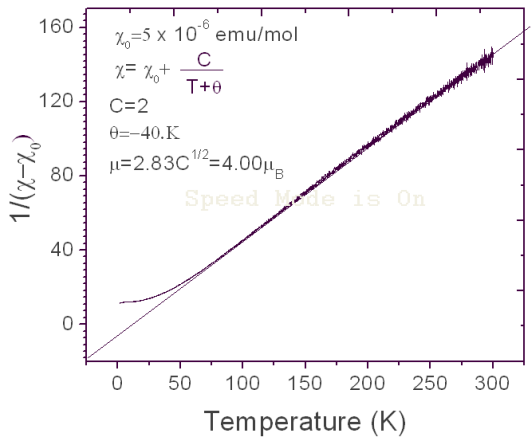}}
\caption{Inverse Magnetic Susceptibility versus Temperature of PrCoIn$_5$ at H=1kOe}
\label{fig:fig2}
\end{figure}
At high temperatures ($T \geq  100K$), $\chi^{-1}$ is linear and
follows Curie-Weiss behavior. We obtained effective moment of
$4\mu_B$. This value is larger than the $3.73\mu_B$ that is expected
for a free Pr$^{3+}$. The higher magnetic moment could be attributed
to an admixture of moments from both Co ions and $Pr^{3+}$. The Co
sub lattice can be magnetized because of f-d exchange as in
$RCo_3$\cite{Gaidukova}.Obviously the moment is dominated by
$Pr^{3+}$ in the $J = 4$ Hund's rule ground state indicating that
the $4f$ electrons are almost localized within the $Pr$ atoms.
Similar results are obtained by others except that our analysis
shows a lower  paramagnetic Curie temperature $\theta \sim-40K$
compared to $-56K$\cite{Nguyen}.

In summary we prepared clean single crystals of $PrCoIn_5$. The
structure is tetragonal it has a c/a$\sim $1.62 slightly  smaller
volume that its counterpart CeCoIn$_5$. Superconductivity is not
observed down to 1K. In addition the behavior of the low temperature
resistivity appears to have originated from a fermi liquid behavior.
Pair breaking phenomenon were observed when CeCoIn$_5$ is alloyed by
the non magnetic La further demonstrating unconventional
superconductivity in this system\cite{Petrovic}. PrCo$_5$ has
slightly lower ($C/a(\sim 1.62$) than that of CeCoIn$_5$ ($C/a(\sim
1.63$). Whether or not this difference has such detrimental effect
on superconductivity is not determined. However the high value of
the magnetic moment and the presumed magnetic order below $T_N \sim
14.5K$ may have led to magnetic pair breaking.

\begin{acknowledgments}
 This work was supported by the NASA MUCERPI2003 under the Grant NNG04GD63G. We are grateful to our collaborators  National High Magnetic Field Laboratory-Los Alamos.

\end{acknowledgments}

\end{document}